\documentclass[%
nofootinbib,
 amsmath,amssymb,
 aps,
floatfix,
twocolumn
]{revtex4-2}

\usepackage{graphicx}
\usepackage{dcolumn}
\usepackage{bm}
\usepackage{color}
\usepackage[dvipsnames]{xcolor}
\usepackage{hyperref}
\hypersetup{colorlinks=true, citecolor=MidnightBlue,linkcolor=CornflowerBlue, urlcolor=CornflowerBlue, linktocpage=true}
\usepackage[section]{placeins}
\usepackage{soul}

\DeclareUnicodeCharacter{2212}{\textendash}

\begin{document}

\title{Separating Astrophysics and Geometry in Black Hole Images}

\author{Guillermo Lara}
\email{jlaradel@sissa.it}
\author{Sebastian H. V\"olkel}
\email{svoelkel@sissa.it}
\author{Enrico Barausse}
\email{barausse@sissa.it}

\affiliation{SISSA - Scuola Internazionale Superiore di Studi Avanzati, via Bonomea 265, 34136 Trieste, Italy and INFN Sezione di Trieste}
\affiliation{IFPU - Institute for Fundamental Physics of the Universe, via Beirut 2, 34014 Trieste, Italy}

\date{\today}

\begin{abstract}
The observation of the shadow of the supermassive black hole M87$^*$ by the Event Horizon Telescope (EHT) is sensitive to the spacetime geometry near the circular photon orbit and beyond, and it thus has the potential to test general relativity in the strong field regime. Obstacles to this program, however, include degeneracies between putative deviations from general relativity and
both the description of the accretion flow
and the uncertainties on 
``calibration parameters'', such as e.g. the 
mass and spin of the black hole.
In this work, we introduce a formalism, based on a principal component analysis, capable of reconstructing the black hole metric (i.e. the ``signal'') in an agnostic way, while subtracting the ``foreground'' due to the uncertainties in the calibration parameters and the modelling of the accretion flow. We apply our technique to simulated mock data for spherically symmetric black holes surrounded by a thick accretion disk. We show that separation of signal and foreground may be possible with next generation EHT-like experiments.
\end{abstract}

\maketitle  

\section{Introduction}\label{introduction}

More than a century after Albert Einstein introduced general relativity (GR) \cite{Einstein:1916vd}, many large scale observational campaigns and theoretical studies are still searching for possible, new insights into gravity. Although GR has so far passed all tests with flying colors, e.g., in recent years the exploration of the dynamical and strong field regime via measurements of black hole (BH) and neutron star mergers  by LIGO and Virgo \cite{PhysRevLett.116.061102,PhysRevLett.119.161101,PhysRevLett.116.221101,Abbott:2020jks}, many of the outstanding open problems in physics involve gravitation and BHs. The latter are among the most extreme objects in nature, and  access to the gravitational waves that they emit, as well as the first radio ``image'' of the shadow of the supermassive BH M87$^*$ by the Event Horizon Telescope (EHT) \cite{eht,Akiyama:2019eap}, allow for testing our theoretical understanding in unprecedented ways. 

One possible way to look for deviations from GR is to test the Kerr hypothesis \cite{Robinson:1975bv,Israel:1967wq,Hawking:1971vc}. The latter states that all astrophysical BHs are uniquely characterized by their mass and angular momentum through the Kerr metric \cite{Kerr:1963ud}. The recent measurement of the shadow of M87$^*$ is therefore particularly interesting to test the Kerr hypothesis and constrain possible deviations from it. Several theoretical works have connected the bright emission ring that is present in the reconstructed image with the impact parameter of the circular photon orbit \cite{Takahashi:2005hy,Johannsen:2010ru,Psaltis:2010ca,Amarilla:2011fx,Loeb:2013lfa,Psaltis:2014mca,Johannsen:2015hib,Psaltis:2015uza,Cunha:2015yba,Cunha:2016wzk,Psaltis:2018xkc,Cunha:2019dwb,Medeiros:2019cde,Younsi:2016azx}. This feature in the image may therefore be used to connect the underlying BH spacetime with the observed image and therefore test the Kerr hypothesis.

Despite this large amount of existing theoretical work on the topic, a lively discussion about the interpretation of BH shadow images is currently ongoing. More precisely, one question is how suited the shadow size alone (as opposed to the full image) is for testing deviations from the Kerr metric. In Psaltis \textit{et al.}~\cite{PhysRevLett.125.141104}, it was claimed that the $17\,\%$ constraint on the shadow size of M87$^{*}$, as reported earlier by the EHT Collaboration in Refs.~\cite{eht,Akiyama:2019eap}, can be used to place constraints beyond the first post-Newtonian order of the BH metric. However, the basis and robustness of this conclusion have been critically discussed since then. Gralla~\cite{Gralla:2020pra} argued that uncertainties in the underlying astrophysics make tests of GR with the current observations not possible. V\"olkel \textit{et al.}~\cite{Volkel:2020xlc} showed that even if some underlying assumptions questioned by Gralla~\cite{Gralla:2020pra} and adopted in Psaltis \textit{et al.}~\cite{PhysRevLett.125.141104} are true, the bounds of Ref.~\cite{PhysRevLett.125.141104} do not hold when higher order post-Newtonian orders are included in the analysis. Nevertheless, V\"olkel \textit{et al.}~\cite{Volkel:2020xlc}, as well as 
Kocherlakota \textit{et al.}~\cite{Kocherlakota:2021dcv}, demonstrate that theory specific tests can be performed and allow for gaining information on the gravitational theory. The role of dimensional coupling constants in alternative theories of gravity has been studied by Glampedakis and Pappas \cite{Glampedakis:2021oie}. 

Since the shadow size, as used in most previous works\footnote{See Nampalliwar \textit{et al.} \cite{Nampalliwar:2021oqr} for a recent work including also the deviations from circularity.}, is only a single number, it represents an immense reduction of the information that is encoded in the full image. For this reason, and because it is based on the connection between the impact parameter 
of the circular photon orbit and the image
brightness, more advanced studies are needed.

Improvements of several aspects of the analysis have already  been partially carried out in recent works. For instance, a multiple ring structure 
seems to appear in the shadow image at high angular resolution, although its observation would require
 measurements beyond current EHT capabilities, see e.g. Refs.~\cite{PhysRevD.102.124004,Broderick:2021ohx, MaciekWielgus:2021peu}. Medeiros \textit{et al.} \cite{Medeiros:2019cde} have also investigated deviations from circularity in non-Kerr spacetimes using a principal component analysis (PCA). Formal questions regarding the uniqueness of shadow images and the underlying BH geometry, from the point of view of geodesics, have been recently studied in Ref.~\cite{Lima:2021las}.

Ideally, however,  an analysis should take into account the whole image and simulate both the astrophysics and the background geometry at the same time. Treating this problem in its full complexity is far beyond of what is currently possible. State-of-the-art calculations combine general relativistic magneto-hydrodynamic (GRMHD) simulations of the matter with ray-tracing codes to construct images --see Ref.~\cite{Akiyama:2019eap, Akiyama:2019fyp}. 
However, even in the context of GR, the large computational cost associated to these calculations allows one to vary (at the same time) at most a few parameters describing the accretion flow physics and the geometry \cite{Akiyama:2019fyp, Vincent:2020dij}. 
Therefore, repeating these simulations for non-Kerr spacetimes is already challenging, even when backreaction on the geometry is neglected --see e.g. Refs.~\cite{Mizuno:2018lxz, Olivares:2020wpo}.
Nevertheless, varying the accretion flow physics and allowing for non-Kerr geometries at the same time has been attempted before, although restricting to particular deformations of the metric~\cite{Broderick_2014}.

In this work, we present a framework 
that allows one to deal with deviations of both
the accretion flow and the spacetime geometry from
standard scenarios, showing 
that under suitable assumptions they can 
be both recovered/constrained from the shadow image.
In more detail, 
 we perform a PCA, which allows us to probe small but very general (i.e. theory ``agnostic'') deviations from the Schwarzschild metric,
 while at the same time reconstructing the astrophysical accretion flow (which 
 we assume to be spherical and with  a simple emissivity profile). The PCA allows for working in a rather general set of basis functions, e.g. Gaussians or power laws. Although our matter description is not as sophisticated as current GRMHD simulations, we show for the first time how the background metric and simple astrophysical scenarios can be disentangled from one another 
 and reconstructed singularly in an \textit{inverse problem for the whole shadow image}. Our work is therefore extensible  to incorporate more realistic models in the future.

Our main findings can be summarized as follows. We provide different mock images produced by our model as hypothetically observed data. These images have been produced by varying the emissivity profile and/or the BH metric. We  have first  demonstrated that the PCA reconstruction allows one to constrain the emissivity profile when the background is assumed to be Schwarzschild. Then we have also demonstrated that general deviations of the metric away from Schwarzschild can be recovered when the emissivity profile is well understood. In the final and most important application, we have demonstrated that when the functional form of the emissivity profile is assumed to be of a simple form, but with unknown parameters, it is possible to recover both the spacetime geometry (the ``signal'') in an agnostic way, as well as the emissivity profile (the ``matter foreground'').

This work is organized as follows. In Sec.~\ref{theory} we summarize  our accretion model and the ray-tracing algorithm. In Sec.~\ref{methods} we outline our numerical method to compute BH images and the PCA. Applications and results are reported in Sec.~\ref{applications_results}. A discussion of our findings is presented in Sec.~\ref{discussion}, while our final conclusions are provided in Sec.~\ref{conclusions}. The metric signature throughout this paper is \(-+++\) and we set \(G = c = 1\).

\section{Theory}\label{theory}

In the following, we will give a brief overview of our modelling of accretion flows (in Sec.~\ref{theory_astro}) and of ray-tracing around supermassive BHs (in Sec.~\ref{theory_ray}).

\subsection{Accretion model}\label{theory_astro}

Being a bright source for terrestrial and space observatories across different parts of the electromagnetic spectrum, M87$^*$ has been extensively observed in a  variety of wavelengths~\cite{EventHorizonTelescope:2021dvx}. The central BH mass can be estimated by modeling the dynamics of nearby gas \cite{Walsh:2013uua} or stars \cite{2011ApJ...729..119G}. Note that these methods give different BH mass values, but the recent EHT measurement of $M_\text{BH} = (6.5 \pm 0.7)\times 10^9 M_\odot$  only agrees with the stellar dynamics measurement \cite{Akiyama:2019eap}. 
The environment of M87\(^{*}\) is believed to be a geometrically thick
and optically thin accretion disk~\cite{Akiyama:2019fyp}, rather than a geometrically thin and optically thick one (such as e.g. the classic Novikov-Thorne model \cite{LesAstresOcclus:1973blho.conf.....D}).
The  photons observed by the EHT at \(1.3\, \mathrm{mm}\) are thought to be produced by synchrotron radiation from the relativistic electrons in the hot accretion plasma~\cite{eht, Akiyama:2019fyp}. Moreover, M87$^*$ exhibits a visible jet, observable at all wavelengths, whose power has been used to disfavor zero-spin models~\cite{Akiyama:2019fyp, Jeter:2020lkw}.

For our purposes, we employ a simplified toy model for the accretion flow. We assume a spherically symmetric and optically thin disk surrounding a spherically symmetric (i.e. non-rotating) BH with metric
\begin{align} \label{eq: metric_ansatz}
    ds^2 & = g_{tt} (r) \, dt^2 + g_{rr} (r)\, dr^2 + r^2 d\theta^2 + r^2 \sin^2 \left(\theta \right) d \phi^2,
\end{align}
 where \(g_{tt}(r)\,g_{rr}(r) = -B(r)^2\), with $B(r)$ being a free function.
The  power radiated by the disk can be characterized by the 
\emph{emissivity} \cite{Lightman:1986},
\begin{align}
    j_\nu(r) &= \dfrac{dE}{dV \, dt\, d\nu},
\end{align}
which we assume to be independent of the  frequency \(\nu\). Motivated by radiatively inefficient accretion flow (RIAF) models \cite{, Broderick:2010kx, Broderick_2014}, we assume that the spatial distribution of the emissivity  is well-described by a power law \(j_\nu(r) \propto  r^{-n}\), with \(n \approx 1\). 

We assume that the BH image is detected 
by a distant observer. 
Hence, the image can be described by the intensity \(I_\nu (b)\), as a function of the impact parameter \(b\) of the null geodesics along which photons propagate. The intensity can be obtained by integrating the radiative transfer equations (here presented in the form of Ref.~\cite{Younsi_2012,RezzollaYounsiBronzwaer:2018lde}),
\begin{align}\label{eq: RTeqs}
    \dfrac{d}{d \lambda} \left(\dfrac{I_{\nu, \text{obs}}}{\nu_\text{obs}^3}\right) & = \dfrac{j_\nu}{\nu^2} e^{-\tau_{\nu, \text{obs}}}, \\
    \dfrac{d \tau_{\nu, \text{obs}}}{d\lambda} & = \alpha_\nu \nu,
\end{align}
where \(\nu = - \kappa_\mu u^\mu\), 
with \(\kappa^\mu\) the photon's linear momentum and
 \(u^\mu\) the 4-velocity of the disk's fluid, is the photon's frequency as measured in the fluid's rest frame.
The subscript ``obs'' denotes quantities in the observer's frame.

These equations are integrated from the matter source to the far away observer along  null geodesics
parametrized by the affine parameter \(\lambda\). Since we make the assumption of a disk that is (perfectly)
optically thin, we choose an absorption coefficient \(\alpha_\nu = 0\), and we also
neglect the Doppler shift due to the motion of the disk's fluid, i.e. we assume 
\(u^{\mu} = (1/\sqrt{-g_{tt}}, 0, 0, 0)\). Furthermore, we assume that  the observer  measures \(I_{\nu, \text{obs}}(b)\) at a single  frequency.\footnote{In practice, the EHT measures \emph{visibilities}, which correspond to the Fourier transform coefficients of the Stokes parameters of the image.} Hereinafter, for simplicity, we will drop the labels \(\nu\) and ``obs'' in both the intensity and the emissivity.

\subsection{Ray-tracing}\label{theory_ray}

The photons contributing to the BH image probe the spacetime along null geodesics. In the vicinity of a Schwarzschild BH, these photons are strongly lensed. For a critical impact parameter \(b_\text{ph} = 3 \sqrt{3} \, M_\text{BH} \), where \(M_\text{BH}\) is the mass of the BH, photons follow unstable circular orbits at a surface called the \emph{photon sphere}. If no matter is present within this surface, this critical value of \(b_\text{ph} \) can be interpreted as the size of the BH \emph{shadow}. \cite{Gralla:2019xty}

In practice, although photons follow trajectories from the source to the observer, \emph{ray-tracing} algorithms often integrate the geodesics and radiative transfer equations in the opposite direction -- see e.g. Ref.~\cite{RezzollaYounsiBronzwaer:2018lde}. In addition, although the geodesics equations are integrable in spherical symmetry (and in the Kerr spacetime), it is convenient to consider the standard second-order equations 
\begin{align}\label{eq: GeodEqs}
    \dfrac{d^2 \kappa^\mu}{d \lambda^2 } + \Gamma^{\mu}_{\rho \sigma} \dfrac{d \kappa^\rho}{d \lambda}\dfrac{d \kappa^\sigma}{d \lambda} & = 0,
\end{align}
where \(\Gamma_{\rho \sigma}^{\mu}\) are the Christoffel symbols.  
The advantages of this choice are that one avoids a special treatment at the turning points of the geodesics, and (as we will describe below) one can linearly perturb these equations in a straightforward way. 

\section{Methods}\label{methods}

In this Section, we outline our PCA method. In more detail, in Sec.~\ref{methods_basis} we introduce the  basis functions on which we decompose
the accretion model and the deviations of the BH metric away from Schwarzschild. In Sec.~\ref{method_linear_likelihood}, 
we present our (linearized)
 likelihood,
 and apply the PCA to it in  Sec.~\ref{method_pca}. The role of  priors is discussed in Sec.~\ref{methods_priors}. 

\subsection{Choice of basis}\label{methods_basis}

A useful way to describe and test the multiple possible modified gravity solutions for BH geometries is to employ physically motivated parametrizations of the metric. In  spherical symmetry, these can take the form of series of \(r^{-n}\) terms, with \(n\neq 0\), like in the post-Newtonian (PN) series, or more complicated expressions, like in the Rezzolla-Zhidenko parametrization, where the metric coefficients are described by Pad\'e approximants \cite{PhysRevD.90.084009, Konoplya:2016jvv}.
One must be careful, however, that any particular parametrization may not be able to describe all possible {\it arbitrary} departures from GR, especially if only few  parameters are left free to vary.\footnote{Note however that the Rezzolla-Zhidenko parametrization has proven very flexible in this respect, as it is capable of reproducing BH geometries in entire classes of theories~\cite{konoplya2020general} with relatively few parameters. However, those theories are not necessarily comprehensive of all possible deviations from GR that one can conceive.}  
For instance, the PN expansion is valid only for mild gravitational fields, c.f. for instance Ref.~\cite{Volkel:2020xlc}, and an infinite number of terms would be needed to describe all possible conceivable BH metrics in the strong field region.

In the following, we therefore allow  the deviations of the metric 
from the Schwarzschild solution to have an arbitrary form. Focusing on  \(g_{tt}(r)\)
and assuming for simplicity a toy model where $B=1$ (and hence $g_{tt}(r)\, g_{rr}(r)=-1$), one can write
\begin{align}\label{eq: gttParam}
    g_{tt}(r) & = g_{tt}^{(0)}(r) + \sum_{i} \alpha^{(tt)}_{i} \delta g_{tt}^{(i)}(r), 
\end{align}
where $\alpha^{(tt)}_{i}$ are free coefficients,  \(g_{tt}^{(0)}(r)\) is the Schwarzschild solution, and the functions \(\delta g_{tt}^{(i)}(r)\) form a suitable \emph{basis} 
on which any smooth 
function defined   on the (positive) real axis can be decomposed. The basis is sometimes referred to also as \emph{frame}. A familiar example of basis functions is given by sines and cosines in Fourier analysis. Another example is provided by the Morlet-Gabor wavelets used in gravitational wave analysis pipelines \cite{Cornish_2021}.
This basis need not be orthogonal. For instance,
in some of the applications described in the following, we choose the basis to consist of Gaussians (centered on different points on the real axis, labelled by a discrete index \(i\), but with a non-vanishing overlap to enforce continuity of the reconstructed function).  

In order to account for uncertainties in the astrophysical model, we can write a similar expression for the emissivity of the disk:
\begin{align}\label{eq: emissivityParam}
        j(r) & = j^{(0)}(r) + \sum_{i} \alpha_{i}^{(J)} \delta j^{(i)} (r),
\end{align}
where $ \alpha_{i}^{(J)}$ are again free coefficients, and \(\delta j^{(i)} (r)\) may be different from the metric basis functions.

\subsection{Linearized Model and Likelihood}\label{method_linear_likelihood}

We approach the inverse problem of reconstructing the spacetime metric from the observed BH image within a Bayesian perspective. The expressions in Eqs.~\eqref{eq: gttParam} and \eqref{eq: emissivityParam} will be the backbone of our model \(I_M(\alpha, b)\) for the BH image. We will then seek to estimate the parameters \(\boldsymbol{\alpha} = \left(\alpha^{(tt)}_{l}, \alpha^{(J)}_{m}\right)\) that best describe the  BH image data \(I_D(b)\). When the posterior probability can be approximated as Gaussian, one can further ``clean'' the reconstructed metric by means of a PCA.

We begin by describing our model for the BH image in more detail. We assume that the image consists of a finite number of data points at locations \(b_1, \dots, b_N\), where the associated intensity  \(\boldsymbol{I}_D = \left(I_{D, 1}, \dots, I_{D, N}\right)\) is measured. We also assume that the data are subject to Gaussian measurement errors, which we assume to be constant and given by $\sigma$. As for 
our model, which we denote by 
\(\boldsymbol{I}_M \left(\boldsymbol{\alpha}\right) = \left(I_M(\boldsymbol{\alpha}, b_1), \dots, I_M(\boldsymbol{\alpha}, b_N)\right)\), we integrate numerically the radiative transfer and geodesic equations [Eqs.~\eqref{eq: RTeqs} and \eqref{eq: GeodEqs}] assuming Eqs.~\eqref{eq: gttParam} and \eqref{eq: emissivityParam}. In particular, in order to render the posterior probability function Gaussian (i.e. quadratic in the parameters $\boldsymbol{\alpha}$) and apply the PCA technique, we linearize Eqs.~\eqref{eq: RTeqs} and \eqref{eq: GeodEqs} in $\boldsymbol{\alpha}$. 
 
Physically, this amounts to assuming that the deviations from Schwarzschild and from our default emissivity model are small. 
(We do not report these linearized equations here as they are cumbersome and not particularly illuminating -- see e.g. Appendix A of Ref.~\cite{Gralla:2012db} for the linear perturbations of the geodesic equations.) In practice, the  14 first-order linearized equations for the variables 
\begin{align}
\{t, r, \phi,\dot t, \dot r, \dot \phi,\delta t, \delta r, \delta \phi, \delta \dot t, \delta \dot r, \delta \dot \phi, I^{(0)}, \delta I\},
\end{align}
where \(\dot X \equiv dX/d\lambda\), are integrated with a custom-made ray-tracing code written in \textsc{C++} and  employing an adaptive stepsize fourth order Runge-Kutta algorithm \cite{Press:1992}. 
Therefore, the model can be written as
\begin{align} \label{eq: IModel}
    \boldsymbol{I}_M \left(\boldsymbol{\alpha}\right)  &= \boldsymbol{I}^{(0)} + \sum_{i = 1}^{M} \alpha_i \delta \boldsymbol{I},
\end{align}
where \(M\) is the total number of \(\alpha_i\) parameters and \(\delta \boldsymbol{I}\) are the image deviations corresponding to each of the individual basis functions.\footnote{Notice that the basis functions (dependent on the coordinate variable \(r\)) are mapped into an image space (dependent on the impact parameter variable \(b\)) by a nonlinear transformation or ``transfer function". In many applications of the PCA (e.g. Ref.~\cite{Pieroni:2020rob}) there is no such mapping.}

In the Bayesian framework, the solution to the inverse problem, up to a normalization factor, is encoded in the posterior probability distribution, given by
\begin{align} \label{eq: BayesTheorem}
    p\left(\boldsymbol{\alpha} \lvert \boldsymbol{I}_D\right) & \propto p\left(\boldsymbol{I}_D \lvert \boldsymbol{\alpha} \right) p\left(\boldsymbol{\alpha}\right),
\end{align}
where the likelihood follows from the assumption of Gaussian measurement errors and is given by
\begin{align} \label{eq: Likelihood}
    \log p\left(\boldsymbol{I}_D \lvert \boldsymbol{\alpha} \right) & = - \dfrac{\chi^2}{2}, 
\end{align}
with
\begin{align}
    \chi^2 & = \dfrac{1}{\sigma^2}\left( \boldsymbol{I}_D - \boldsymbol{I}_M \left(\boldsymbol{\alpha}\right) \right)^T \left( \boldsymbol{I}_D - \boldsymbol{I}_M \left(\boldsymbol{\alpha}\right)\right).\label{likelihood}
\end{align}
In addition, as explained below, we will assume Gaussian priors \(p \left(\boldsymbol{\alpha}\right)\). This choice, coupled with the likelihood Eq.~\eqref{likelihood}, yields Gaussian posteriors, which are suitable for PCA.

\subsection{Principal Component Analysis}\label{method_pca}

Since the model is linear in the parameters and the posterior probability is Gaussian, the maximum of the latter (i.e. the ``most likely'' parameters \(\boldsymbol{\alpha}^\star\)) can be obtained by solving a (possibly degenerate) linear system of equations of the form \(\boldsymbol{F} \boldsymbol{\alpha}+\boldsymbol{q} = 0\), where the \(M\times M\) matrix \(\boldsymbol{F}\) (defined below) and the \(M\)-vector \(\boldsymbol{q}\) are computed numerically with our ray-tracing code. (This is the most computationally expensive part of the framework since it involves computing a perturbed image for every basis function.)
The errors associated with the parameters are in general correlated and are encoded in the Fisher (Hessian) matrix 
\begin{align} \label{eq: FisherMatrix}
    F_{lm} & = - \dfrac{1}{2}\dfrac{\partial^2}{\partial \alpha_l \partial \alpha_m} \log   p\left(\boldsymbol{\alpha} \lvert \boldsymbol{I}_D\right),
\end{align}
which, with our assumptions, becomes a constant matrix.
Linear combinations of the parameters corresponding to the Fisher matrix orthonormal eigenvectors \(\boldsymbol{e}^{(i)}\), however, have uncorrelated errors \(\sigma^{(i)} = 1/\sqrt{2\lambda^{(i)}} \), where \(\lambda^{(i)}\) are the corresponding eigenvalues. 

The main idea behind the PCA is to
clean the reconstruction of the 
model by keeping only the ``largest'' coefficients 
\(\beta_i = \boldsymbol{\alpha}^\star \cdot \boldsymbol{e}^{(i)}\).
This is akin to the procedure of cleaning a time series from noise by performing a Fourier transform, and then keeping only the
Fourier terms with coefficients significantly different from zero.

More precisely, we prescribe the selection criterion \cite{Pieroni:2020rob}
\begin{align}
    \left \lvert \beta_i \right \rvert \geq N_\text{th} \sigma^{(i)},
\end{align}
whereby we only retain the coefficients \(\beta_i\) that are inconsistent with zero at \(N_\text{th}\)-th sigma level,
with $N_\text{th}\approx 1$-3. 
The final estimate of the parameters is then  given by
\begin{align}
    \boldsymbol{\alpha}^\text{PCA} & \equiv \sum_{\left \lvert \beta_i \right \rvert \geq N_\text{th} \sigma^{(i)}} \left(\beta_i\ \pm \sigma^{(i)}\right) \boldsymbol{e}^{(i)}.
\end{align}
Explicitly, the reconstructed metric and emissivity are then given by
\begin{align}
    g_{tt}^\text{PCA}(r)& \equiv g_{tt}^{(0)}(r) + \sum_{\left \lvert \beta_i \right \rvert \geq N_\text{th} \sigma^{(i)}} \left(\beta_i\ \pm \sigma^{(i)}\right)\eta^{(i),(tt)}(r) ,\\
        j^\text{PCA}(r)& \equiv j^{(0)}(r) + \sum_{\left \lvert \beta_i \right \rvert \geq N_\text{th} \sigma^{(i)}} \left(\beta_i\ \pm \sigma^{(i)}\right)\eta^{(i),(J)}(r),
\end{align}
where
\begin{align}
    \eta^{(i),(tt)}(r) &\equiv \sum_{k} e^{(i),(tt)}_k \delta g_{tt}^{(k)} (r),\\
    \eta^{(i),(J)}(r) &\equiv \sum_{k} e^{(i),(J)}_k \delta j^{(k)} (r) ,
\end{align}
are the corresponding eigenfunctions. Here we have separated the eigenvectors \( \boldsymbol{e}^{(i)} = \left(e^{(i),(tt)}_l, e^{(i),(J)}_m\right) \) according to their corresponding metric and emissivity indices.
Note that because the coefficients
$\beta_i$ are uncorrelated Gaussian variables, the errors on the reconstructed metric and emissivity at  a given location can be computed by propagating the errors $\sigma^{(i)}$ in quadrature. These errors will be reported in Sec.~\ref{applications_results}.

\subsection{Priors}\label{methods_priors}

The use of priors in the reconstruction  can enhance it by mitigating some of the degeneracies that may be present. First, the linear system needed to solve for \(\boldsymbol{\alpha}^\star\), and thus  to obtain \(\boldsymbol{\alpha}^\text{PCA}\), is in general degenerate. One way to tame this issue is by conditioning the Fisher matrix \(F_{lm}\), i.e. by replacing \(F_{lm} \to F_{lm} + \epsilon \, \delta_{lm}\), where \(\delta_{lm}\) is the Kronecker delta and \(\epsilon\) is suitably small \cite{Pieroni:2020rob}. From the Bayesian perspective, this can be interpreted as prescribing loose Gaussian priors for \(\boldsymbol{\alpha}\) centered at zero. At least in the case of the metric, this is in line with the theoretical expectation that the parameters \(\alpha_{i}^{(tt)}\), which describe the deviations from GR, should be small.

Proper priors on the coefficients of neighboring basis functions can also be used to enforce continuity \cite{Silva:2017uqg}. More importantly, from a physical point of view, priors can be used to enforce the expectation that the metric deviations are well constrained at large distances, where the Newtonian limit of GR (but not necessarily its 1PN dynamics~\cite{Volkel:2020xlc}) should be recovered. For the Gaussian basis functions uniformly distributed in radius that we will use below, we therefore include a prior of the form
\begin{align}\label{eq: prior_large_distance}
p(\alpha) \propto \exp \left(-\left(\frac{r_{(i)}}{M_\text{BH}}\right)^n\sum_{i }^{}\frac{\alpha_{i}^{(tt)\,2}}{2\sigma^2_r}\right),
\end{align}
with \(n = 4\), constant \(\sigma_r\) and \(r_{(i)}\) the location of the Gaussian basis function associated to \(\alpha_i^{(tt)}\). Equivalently, this corresponds to the expectation that the deviations from GR enter at 1PN order or higher~\cite{Volkel:2020xlc}.
This prior has also the advantage that it stabilizes the reconstruction against fluctuations from different noise realizations. 
Finally, the choice of the reference functions \(g_{tt}^{(0)}(r)\) and \(j^{(0)}(r)\) constitutes an additional ``theoretical prior''. 

\section{Applications and Results}\label{applications_results}

We apply our framework to the following cases. First, in Sec.~\ref{app_astro} we assume that the spacetime geometry is known and only the accretion flow, described by the emissivity, needs to be reconstructed. Second, in Sec.~\ref{app_geometry} we consider the opposite case in which the emissivity is assumed to be known, but the spacetime geometry needs to be reconstructed. Finally, in Sec.~\ref{app_astro_geometry} we let both the spacetime geometry and the accretion flow model undetermined  at the same time. In order to directly compare the results obtained in these three ways, we show the corresponding reconstructions in the same figures at the end of this Section.
Since we do not realistically model the details of the noise in the simulated observations, in the following we will focus on injections in the noiseless approximation. This for instance standard when designing gravitational wave data analysis pipelines~\cite{Rodriguez:2013oaa}.
However, we have checked that the results are similar for explicit realizations of the Gaussian noise.

In the following, we also introduce the scales \(j_{*}\) and \(I_{*}\), with units 
\( [j_{*}] = [\text{energy}] \, [\text{length}]^{-3} \) and
\([I_{*}] = [\text{length}] \, [j_{*}]\),
as a normalizing scales for the emissivity and observed intensity. 

\subsection{Reconstructing the Accretion Flow}\label{app_astro}

We fix the reference metric function 
to be that of the Schwarzschild geometry,  \(g^{(0)}_{tt}(r) = - ( 1 - 2 M_\text{BH} / r ) \), without allowing for any deviations from it --i.e. only the astrophysical parameters \(\alpha^{(J)}_{i}\) are allowed to vary. As for the reference emissivity, we choose it to be \( j^{(0)}(r) = 0\) (i.e. we assume no prior knowledge of the emissivity). 
We produce the injection (i.e. the data) \(\boldsymbol{I}_D\) from
\begin{align} \label{eq: Injection1}
    g_{tt,D}(r) &= g^{(0)}_{tt} (r) , &
    j_{D}(r) &= 2  j_{*}\dfrac{M_\text{BH}}{r}.
\end{align}
We assume an optimistic resolution of \(0.15\, M_\text{BH}\) (potentially achievable with future space-based interferometers \cite{Johnson_2020}). More precisely, we generate
 \(N = 100\) data points uniformly distributed in the interval of impact parameters \([0, 15\,M_\text{BH}]\), 
and assume an (uncorrelated) measurement error of \(\sigma = 0.1 \, I_{*}\). 
The basis is composed of 161 (unnormalized) Gaussians uniformly distributed in the interval of radii \([0, 100\, M_\text{BH}]\), with root mean square (RMS) width of \(1 \, M_\text{BH}\). 
We condition the Fisher matrix with \(\epsilon = 1\), which can also be interpreted as a Gaussian prior, as discussed in Sec.~\ref{methods_priors}.

In the upper panel of Fig.~\ref{fig_app_1}, in blue, we show the intensity profile for the injection Eq.~\eqref{eq: Injection1}, as a function of the impact parameter. 
One can clearly observe the effect of the photon sphere at \(b \approx 5.2 \, M_\text{BH}\). 
In Fig.~\ref{fig_app_3}, with blue contours, we show the \(2\sigma\) bands of the reconstructed emissivity profile for a PCA criterion parameter \(N_\text{th} = 1\). We observe good agreement with the injection  (gray dashed lines) for  most radii. The reconstruction, however, does not accurately reproduce the emissivity profile near the BH horizon at \(r = 2 \, M_\text{BH}\). This disagreement is most likely due to the effect of gravitational redshift, which suppresses the intensity of the near-horizon Gaussian components. The widening of the reconstruction bands
at \(r \approx 14 \, M_\text{BH}\) occurs instead because the image data were provided in a finite interval --i.e. it is an ``edge effect".

\subsection{Reconstructing the Geometry}\label{app_geometry}

In the following, we present the results of our framework when the accretion flow is assumed to be known and only the background metric is reconstructed --i.e. when only the parameters \(\alpha^{(tt)}_i\) are allowed to vary. 

\begin{figure}[]
\includegraphics[width=1.0\linewidth]{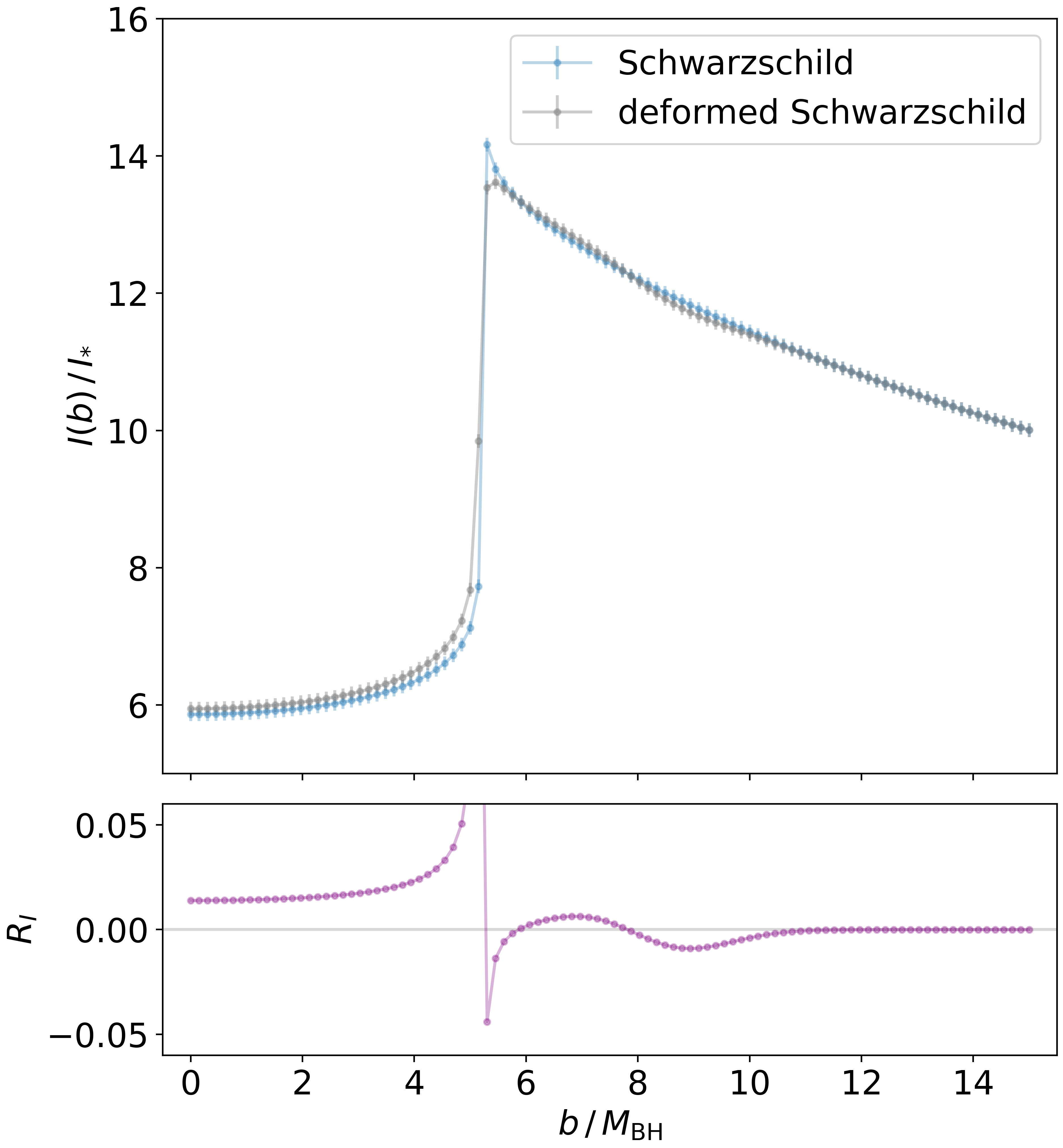}
\caption{Upper panel: Computed intensity profiles $I(b)$  used as ``observations'', as a function of the impact parameter $b$ for the Schwarzschild metric (blue) and the deformed Schwarzschild metric (gray), with corresponding ``measurement'' error bars  of \(\pm 0.1 \, I_{*}\). The scale \(I_{*}\) is used to normalize the intensity profile, and its units are given in the main text. The Schwarzschild intensity profile is used as input data for the example of Sec.~\ref{app_astro}, while the deformed Schwarzschild one is used in the examples of Secs.~\ref{app_geometry} and \ref{app_astro_geometry}. 
Lower panel: Relative difference \( R_I =   \left(I_\text{Sch} -I_\text{def-Sch}  \right)/I_\text{Sch}  \) of the images in the upper panel. \label{fig_app_1}}
\end{figure}

For this example, the injection includes a deviation from GR in the metric, which is given by
\begin{align} \label{eq: Injection2}
    g_{tt,D}(r) &= g^{(0)}_{tt} (r) 
     - 0.02 \, \exp\left[-\dfrac{(r - 7\,M_\text{BH} )^2 }{ M_\text{BH}^2}\right] + \nonumber\\
     & \qquad - 0.8 \, \left(\dfrac{M_\text{BH}}{r}\right)^{3}   , \nonumber\\
    j_{D}(r) &= j^{(0)}(r) =  2 j_{*}  \dfrac{ M_\text{BH}}{r},
\end{align}
where
\(g^{(0)}_{tt}(r) = - ( 1 - 2 M_\text{BH} / r ) \). 
For the sake of simplicity, we also set the reference function for the emissivity to be the same as the injection --although we could have chosen a slightly different profile without strongly affecting our results. 
The resolution and measurement errors are the same as in the previous example. 
We choose a basis  of 161 Gaussians uniformly distributed in the interval \([0, 25 \, M_\text{BH}]\) and with RMS width of \(0.5 \, M_\text{BH}\), supplemented by additional power law components \(r^{-n}\), with \(n = 1, 2, \dots, 15\). To improve the conditioning of the Fisher matrix, we also 
rescale the coefficients \(\boldsymbol{\alpha}\) so that the different entries in that matrix are roughly of the same order of magnitude. 
The  \(r^{-1}\) power law accounts for the uncertainty in our knowledge of the mass, for which we assume a small prior measurement error of \(\sigma_{M_\text{BH}} = 0.01\,M_\text{BH}\). 
Finally, we take \(\epsilon = 10^{-3}\) for the conditioning of the Fisher matrix, and \(\sigma^2_{r} = 10^5\)  for the large distance prior of Eq.~\eqref{eq: prior_large_distance}.

The intensity profile  used as data and produced from Eq.~\eqref{eq: Injection2} is shown in the upper panel of Fig.~\ref{fig_app_1}, in gray. 
Here, one can observe the breakdown of the linear approximation
to the full geodesics equation 
at \(b\approx 5.2 \, M_\text{BH}\), where the linear approximation fails to reproduce the shift in the photon sphere projection away from the Schwarzschild solution, shown in blue. We will not worry about this spurious effect for the moment (as it concerns only few data points) and postpone a more thorough  discussion of it to Sec.~\ref{discussion_comparison}. 

Let us also briefly comment about the features of the resulting image. In the lower panel of Fig.~\ref{fig_app_1}, we show the relative difference between the images of the upper panel of Fig.~\ref{fig_app_1}. Since the injected metric deformation [Eq.~\eqref{eq: Injection2}] has a local minimum around \(r\approx 7\,M_\mathrm{BH}\), one could naively expect that the absolute value of the relative difference should also have a local maximum there. However, a more involved structure is evident. The appearance of a hill and a trough in the intensity is mostly due to the derivatives of the metric deviation from Schwarzschild becoming larger and dominant at the location of the bump -- indeed, this pattern resembles the shape of the derivative of a Gaussian function.

In Fig.~\ref{fig_app_4}, in blue, we show the 2\(\sigma\) reconstruction contours for the deviation of the metric from Schwarzschild. As before, we observe generally good agreement with the injection (gray dashed lines) for the \(N_\text{th} = 1\) PCA criterion. The reconstructed bump is clearly distinguishable from the power law component. The narrower error bands near the photon sphere at \(r \approx 3 \, M_\text{BH}\) indicate greater sensitivity of the method to features in that region. 
As in the previous example,  the goodness of the reconstruction deteriorates near the BH horizon (\(r \approx 2 \, M_\text{BH}\)) presumably due to the effect of gravitational redshift. 
Finally, we observe a slight oscillatory behavior around \(r \approx 11 \, M_\text{BH}\). This happens because many of the eigenvectors (and consequently the eigenfunctions) of the Fisher matrix present oscillatory features. By including more components (i.e. lower \(N_\text{th}\) or smaller  measurement error), these oscillatory components interfere and cancel out to give a better overall reconstruction. 

\subsection{Reconstructing Astrophysics and Geometry}\label{app_astro_geometry}

Finally, we will now present the most interesting results of our framework. These correspond to cases in which both the spacetime and accretion flow are 
reconstructed
at the same time --i.e. all of the parameters \(\boldsymbol{\alpha}\) are allowed to vary. 
Since both aspects play an important role in the details of BH images, one might argue that it is not possible to constrain deviations in the metric when the exact details of the astrophysical model are not known --see for example Gralla~\cite{Gralla:2020pra}. 
In this section, we will show that if our ``theoretical priors" are strong enough, it is in principle possible to disentangle the two types of contributions. Such strong priors can be achieved by sufficient understanding of the underlying astrophysics and by reasonable constraints on the long distance behavior of the metric. 

For this example, the injection and reference functions are the same as in Sec.~\ref{app_geometry}. 
In this case, as expected, we find that there are degeneracies between the metric and the emissivity, when one considers a Gaussian basis for both functions. Indeed, one generically obtains degenerate reconstructions -- i.e. profiles for the metric and emissivity that reproduce the data but do not correspond to the injection. Since these are ``failed'' reconstructions, we do not present them here.

In order to tighten our priors and mitigate this degeneracy, we assume that the radial distribution of the emissivity is well described by a power law (as in RIAF models), and thus allow only the amplitude and exponent to vary. Moreover, we linearize the emissivity in the deviation of the exponent away from the default reference value [which is set as in Eq.~\eqref{eq: Injection2} for this example], so as to reduce to Eq.~\eqref{eq: emissivityParam}. 
We take the metric basis and priors to be the same as in the previous example, 
and perform the reconstruction with a PCA criterion \(N_\text{th} = 1\) as before. 

\begin{figure}[]
\centering
\includegraphics[width=1.0\linewidth]{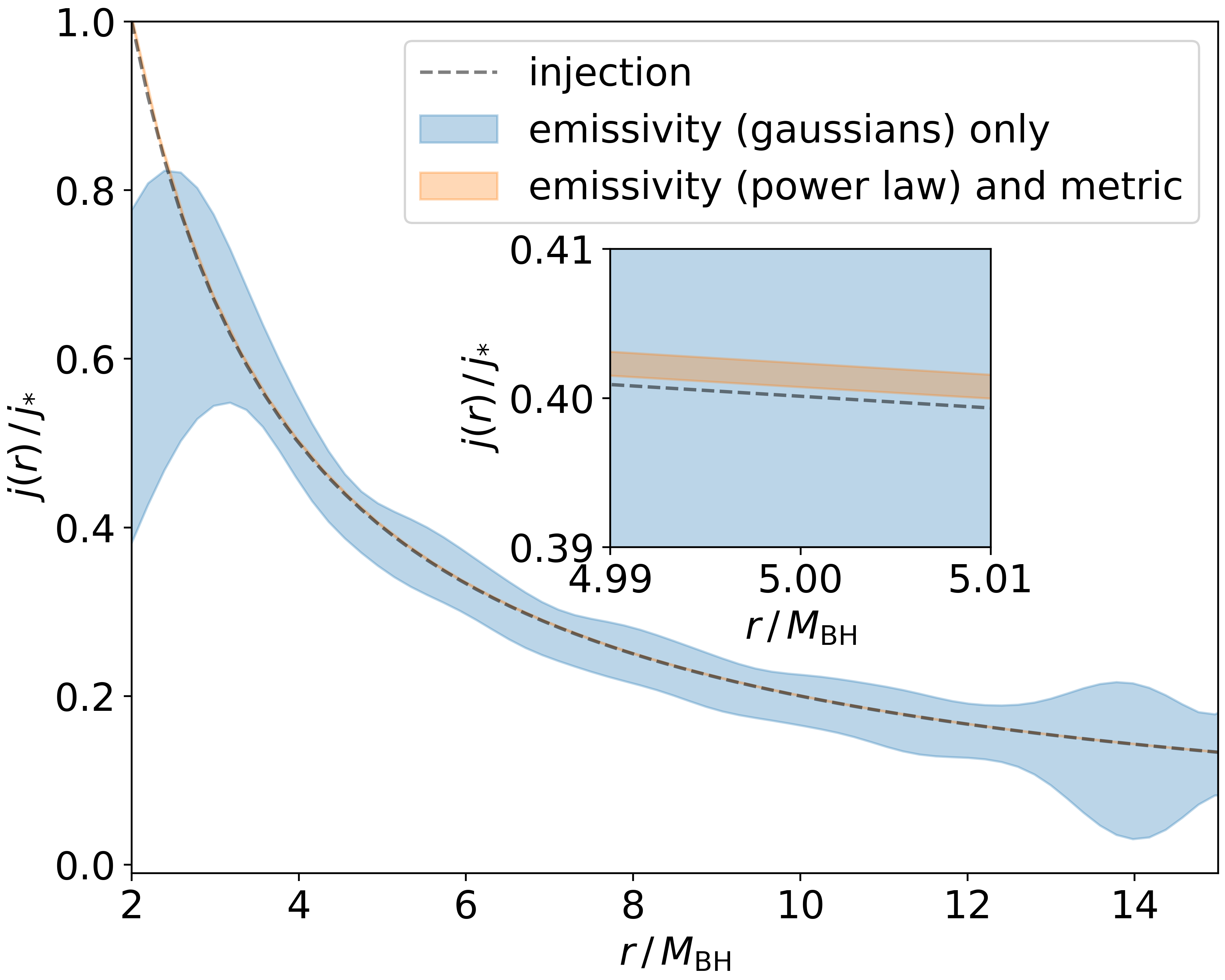}
\caption{Individual reconstruction of the emissivity
when the metric is fixed (blue), and joint reconstruction (orange) when it is constrained to be a power law, corresponding respectively to the examples of Sec.~\ref{app_astro} and \ref{app_astro_geometry}. The scale \(j_{*}\) is used to normalize the emissivity and its units are given in the main text. } \label{fig_app_3}
\end{figure}

In Fig.~\ref{fig_app_3}, we show the 2\(\sigma\) reconstruction contours of the emissivity in orange. 
Because of the tight ``theoretical priors'' that we
assumed on the functional form of the
the emissivity, the
reconstruction is much better than the ``agnostic'' one. 
However, upon closer inspection, as can be seen from the inset, a slight bias is noticeable in the reconstruction. Part of it is due to the missing information weeded out by the PCA criterion.
We have verified this by comparing reconstructions with different values of \(N_\text{th}\). 

\begin{figure}[]
\centering
\includegraphics[width=1.0\linewidth]{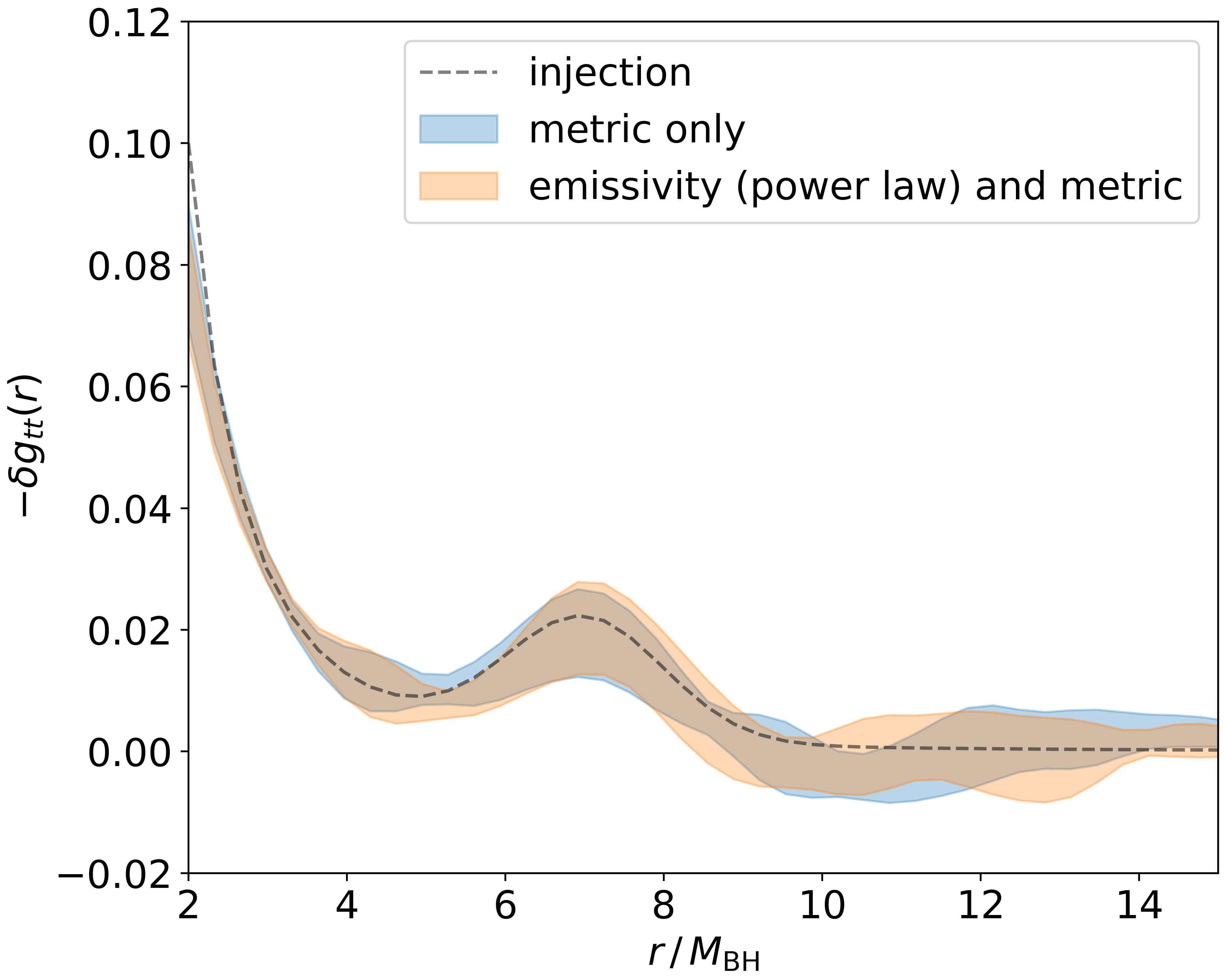}
\caption{Individual reconstruction of the metric deviation from  Schwarzschild, \(\delta g_{tt} (r) = g^{}_{tt} (r) - g^{(0)}_{tt} (r) \),  when the emissivity is fixed  (blue) and joint reconstruction (orange), corresponding respectively to the examples of Sec.~\ref{app_astro} and \ref{app_astro_geometry}.}
\label{fig_app_4}
\end{figure}

In Fig.~\ref{fig_app_4}, one can observe that the reconstruction of the metric deviation (in orange) recovers the injection as in the previous example. It also exhibits the narrower contour bands near \(r \approx 3 \, M_\text{BH}\), and the slightly inaccurate reconstruction near the BH horizon at \( r \approx 2 \, M_\text{BH}\), like before.

This example illustrates that a good, but not necessarily complete, understanding of the astrophysics may be sufficient to extract possible deviations from GR from the BH image. Although constraining the emissivity to follow a particular parametrization -- here a power law -- may seem rather restrictive, this is not necessarily the only way to mitigate the degeneracy. 
Alternative possibilities may include judiciously prescribing the priors of the astrophysical parameters --in the spirit of the metric prior of Eq.~\eqref{eq: prior_large_distance}--, or reducing the number of parameters in the metric by employing a particular ansatz -- e.g. the Rezzolla-Zhidenko parametrization \cite{PhysRevD.90.084009}. 
This second possibility, however, comes at the cost of being less general, however. We leave these possibilities to future extensions of this work. 

\section{Discussion}\label{discussion}

In this Section, we briefly comment about the relation between our framework and other existing approaches 
(Sec.~\ref{discussion_comparison}). Some limitations and possible extensions of this work are discussed in Sec.~\ref{discussion_limitations_and_extensions}. Finally, possible connections to the inverse problem in gravitational wave astronomy are mentioned in Sec.~\ref{discussion_gw}. 

\subsection{Comparison to Other Approaches}\label{discussion_comparison}

Many previous works on modified gravity constraints using the EHT measurements utilize only the size of the critical curve corresponding to the projection at infinity of the photon sphere
\cite{Takahashi:2005hy,Johannsen:2010ru,Psaltis:2010ca,Amarilla:2011fx,Loeb:2013lfa,Psaltis:2014mca,Johannsen:2015hib,Psaltis:2015uza,Cunha:2015yba,Cunha:2016wzk,Younsi:2016azx,Psaltis:2018xkc,Cunha:2019dwb,Medeiros:2019cde}. 
This is justified, because it has been argued that the size of the bright ring is a robust feature in the BH image \cite{Akiyama:2019fyp,PhysRevLett.125.141104,Kocherlakota:2021dcv}. 
However, whether the approach is to constrain the charges and coupling constants of particular solutions in modified theories of gravity (e.g. ``exotic'' Reissner–Nordstr\"om solutions \cite{Garfinkle:1990qj, Gibbons:1987ps}), or to constrain the parameters in phenomenological parametrizations of the metric (e.g. the Rezzolla-Zhidenko parametrization \cite{PhysRevD.90.084009}), the bounds that are obtained are generically degenerate and loose when many parameters are at play \cite{Volkel:2020xlc}. 
This is not surprising since the shadow size amounts to the measurement of a single number. 
Indeed, if our PCA framework is applied  to the shadow size alone (see Appendix~\ref{sec: PCAShadow}), the only combination of parameters that can be constrained [c.f. Eq.~(11) of Ref.~\cite{Volkel:2020xlc}] is
\begin{align} \label{eq: PCAShadowEq}
    \sum_{i = 1}^{M} \alpha_i \delta g_{tt}^{(i)} \left(3\, M_\text{BH}\right).
\end{align}

In this work, we have on the one hand exploited the whole intensity profile of the EHT BH image (although in a toy model and with mock data). On the other hand, to
allow for generic deviations away from GR BHs, we have parametrized the metric in terms of a completely general superposition of basis functions. We have then performed a PCA analysis to obtain a smooth metric reconstruction, filtering out the components that cannot be significantly constrained. 
This technique has been applied before to describe the shape of the shadow of spinning BHs in non-Kerr metrics in Ref.~\cite{Medeiros:2019cde}. However, unlike in previous applications, here it is used to  reconstruct the spacetime geometry  and the disk's emissivity using the whole intensity profile of the BH image.

\subsection{Limitations and possible extensions}\label{discussion_limitations_and_extensions}

Although astrophysical BHs are believed to be simple (i.e. describable only by their mass and angular momentum \cite{Israel:1967wq, Hawking:1971vc, Carter:1971zc, Robinson:1975bv}), the modelling of their environments is fairly complex. In this work, we have made a large number of simplifications (no spin, static and spherical accretion flow, etc.). Nevertheless, the inclusion of many of these environmental aspects is conceptually straightforward, although it would lead to computationally more intensive analyses. 

We have made further simplifications regarding the observational data. In particular, we have assumed that the intensity profile is measured directly and that the associated errors are Gaussian and uncorrelated. The inclusion of \emph{visibilities} -- the observables that the EHT actually measures and which correspond to the Fourier transform coefficients of the image -- is also possible in future extensions of this work. 
However, one would have to 
 deal with additional errors due to the incomplete observational coverage of the visibility parameter space -- see e.g. Fig.~2 of Ref.~\cite{eht}.

Regarding the actual calculation of the BH image, we have  linearized the radiative transfer and geodesic equations [Eqs.~\eqref{eq: GeodEqs} and \eqref{eq: RTeqs}]. This step was essential in order to make the likelihood a Gaussian function and to allow us to apply the PCA technique (which relies on the Fisher matrix being constant). For the case of the metric, this is justified by the expectation that deviations from the GR solution are small. Similarly, the linearization
is reasonable for the
accretion parameters, provided that one has good prior knowledge of the astrophysical model. 

Even in this case, however, some non-perturbative effects  may be important. In particular, the linearization does not capture the shift in the position of the photon sphere or the BH horizon. Future work in this direction may include integration of the full non-linear equations and reconstruction with Markov Chain Monte Carlo (MCMC) techniques, along the line of Refs.~\cite{Broderick_2014, THEMIS:2020eky}. 

Finally, we also recall that we have considered a metric ansatz [Eq.~\eqref{eq: metric_ansatz}] satisfying \(g_{tt}(r)\,g_{rr}(r) = -1 \). It is well known that this condition does not hold in general \cite{Jacobson_2007}. Moreover, it is expected that BHs have spin -- e.g. to power the jet seen in M87\(^{*}\) \cite{Akiyama:2019fyp, Jeter:2020lkw}. 
Therefore, a more rigorous analysis should necessarily include more than one independent metric function, leading to further degeneracy between them. 
A first step in this direction would be to consider the slowly rotating case, by assuming a fixed form of the mixed terms in the metric, i.e. \(g_{t\phi} (r) = g_{\phi t} (r) = - M_\text{BH}^2 a / r \), and linearizing Eqs.~\eqref{eq: RTeqs} and \eqref{eq: GeodEqs} in the dimensionless spin parameter \(a\). In this case (even if the accretion flow is spherical) the radial symmetry of the BH image is broken and we expect that the angle of observation will become important in our ability to extract \(a\) with our PCA approach.
In the most general case, once rotation and realistic accretion are considered, we expect that a good modelling of the astrophysical component, as well as appropriate priors that reflect our knowledge in the weak gravity regime, will continue to be key in allowing to extract possible deviations of the spacetime geometry. Moreover, although more information in the BH image could potentially be extracted, additional strategies may be needed handle degeneracy in the parameter space. We leave this more challenging setup for future work.

\subsection{Possible Connections to Gravitational Waves}\label{discussion_gw}

It is well known in the literature that BH \textit{quasi-normal modes} in the eikonal approximation are closely related to the impact parameter of the BH shadow \cite{PhysRevD.30.295,Yang:2012he}. An interesting question is therefore whether  there is a gravitational wave analog to the BH image. 
Unlike the image, which depends strongly on the surrounding matter, gravitational waves couple only very weakly to the environment, which would allow for a much cleaner test of BH geometries.

One possibility would be to use the full, infinite set of BH quasi-normal modes \cite{Kokkotas:1999bd,Nollert_1999,Berti_2009}. However, the  eikonal approximation only probes the spacetime near the maximum of the potential. Moreover, 
in general one cannot expect that the inverse spectrum problem is uniquely solvable, e.g., in GR axial and polar perturbations are isospectral,  but the underlying Regge-Wheeler and Zerilli potentials are not the same (and other isospectral, and thus equivalent, potentials can be built with suitable transformations \cite{Chandrasekhar:1985kt,Glampedakis:2017rar}).
While degeneracies remain, however, it is possible to put some constraints on the BH metric or the potential for quasi-normal modes,
given a finite set of quasi-normal mode measurements (see e.g.~\cite{paper8,paper10}). 

Closely related to the quasi-normal mode spectrum 
are the \textit{transmission and reflection coefficients}, which describe the frequency dependent wave propagation in the spacetime. These coefficients are computed over the whole exterior BH spacetime. 
In the Wentzel–Kramers–Brillouin (WKB) approximation \cite{1978amms.book.....B} this can be done via integral equations, which one can attempt to invert to constrain the properties of the potentials \cite{Cole:1978zz,Lazenby:1980se,doi:10.1119/1.2190683}. This can also be connected to Hawking radiation \cite{Volkel:2019ahb}. Like for the quasi-normal modes, however, the problem is not uniquely invertible.

\section{Conclusions}\label{conclusions}

In this work, we have demonstrated that BH spacetimes and simple accretion models can be constrained at the same time from BH images. The general inverse problem of BH imaging involves the reconstruction of the astrophysical properties of the accretion disk, as well as possible deviations of the spacetime geometry from GR. Our analysis is a proof of principle, because it is only valid (strictly speaking) for spherically symmetric spacetimes and it makes considerable simplifications for the accretion model and for the mock data. It is clear that realistic modelling of astrophysical processes, as well as of the EHT data analysis pipeline, exceed what can be described by our simple current framework. However, the detailed study of a simplified toy models is still of great value and interest, since it allows one to understand the fundamental aspects of the problem. This is also well demonstrated by the ongoing discussion about the interpretation of the bright emission ring observed by the EHT collaboration, and namely about whether that can be robustly identified with the impact parameter of the photon orbit, and thus be used to  test  the Kerr hypothesis \cite{PhysRevLett.125.141104,Gralla:2020pra,Kocherlakota:2021dcv,Volkel:2020xlc,Glampedakis:2021oie}.

In summary, our framework addresses the fundamental question of whether astrophysical uncertainties and degeneracies jeopardize tests of the Kerr hypothesis with EHT BH images.
 By deriving a linear model that describes the BH geometry in terms of a  general superposition of basis functions (Gaussians and power laws), we have demonstrated that a PCA technique
allows for reconstructing the spacetime geometry {\it and} the accretion model {\it simultaneously}, provided that sufficient theoretical priors are available on the latter. 

\acknowledgments
We thank Avery E. Broderick and Boris Georgiev for useful comments on accretion disk models in the earlier stages of this work, as well as Luciano Rezzolla and Maciek Wielgus for insightful comments on this manuscript. All authors acknowledge financial support provided under the European Union's H2020 ERC Consolidator Grant ``GRavity from Astrophysical to Microscopic Scales'' grant agreement no. GRAMS-815673. This work was supported by the EU Horizon 2020 Research and Innovation Programme under the Marie Skłodowska-Curie Grant Agreement No. 101007855.

\appendix

\section{Black hole shadow with PCA} \label{sec: PCAShadow}

In this appendix we apply the PCA framework to the shadow size of a non-rotating BH and conclude that only one combination of parameters can be meaningfully constrained. We will ignore the effect of spin as in Ref.~\cite{Kocherlakota:2021dcv}.

The impact parameter \(b_\text{ph}\) and radial location \(r_\text{ph}\) of the photon sphere are obtained by solving the system \(V\left(r_\text{ph}, b_\text{ph}\right) = \partial V\left(r_\text{ph}, b_\text{ph}\right)/\partial r = 0\), where \(V(r, b)\) is the effective potential of null rays in spherical symmetry. Remarkably, \(V(r)\) is only sensitive to the \(g_{tt}(r)\) metric function, and therefore, no constraints can be placed on an independent function \(g_{rr}(r)\). Explicitly, this system becomes \cite{Volkel:2020xlc} 
\begin{align}\label{eq: system2}
    r_\text{ph} &= \dfrac{2 g_{tt}(r_\text{ph})}{g'_{tt}(r_\text{ph})} , &
    b_\text{ph}^2 & = - \dfrac{4 g_{tt}(r_\text{ph})}{\left[g'_{tt}(r_\text{ph})\right]^2}    .
\end{align}
For the Schwarzschild metric, these equations can be easily solved to obtain \(r_\text{ph}^\text{Sch} = 3 \, M_\text{BH}\) and \(b_\text{ph}^\text{Sch} = 3\sqrt{3} \, M_\text{BH} \).

A deformed Schwarzschild metric introduces deviations on the photon sphere radius \(\delta r_\text{ph}\) and impact parameter \(\delta b_\text{ph}\). By linearizing Eqs.~\eqref{eq: system2}, they are found to be
\begin{align}\label{eq: deltarb}
    \delta r_\text{ph} &= -\frac{3}{8} \left[3 r_s^2 \,\delta g_{tt} '\left(r_\text{ph}^\text{Sch}\right)-4 r_s \,\delta g_{tt} \left(r_\text{ph}^\text{Sch}\right)\right], \nonumber\\
    \delta b_\text{ph} &= \frac{9}{4} \sqrt{3}\, r_s \,\delta g_{tt} \left(r_\text{ph}^\text{Sch}\right),
\end{align}
where \(r_s = 2 \, M_\text{BH}\) and where we can describe \(\delta g_{tt}(r)\) as a sum of \(M\) basis terms with parameters \(\alpha_i\) as in Eq.~\eqref{eq: gttParam} in the main text. Our model for the impact parameter -- the only observable-- will then be \(b_M\left(\boldsymbol{\alpha}\right) = b_\text{ph}^\text{Sch}  + \delta b_\text{ph}\left(\boldsymbol{\alpha}\right)\).

We compare with a measurement \(b_\text{ph}^\text{Sch} \pm \sigma \), with error \(\sigma \approx 0.17 \, b_\text{ph}^\text{Sch}\) \cite{PhysRevLett.125.141104}, by writing the likelihood
\begin{align}
    \log p( \boldsymbol{\alpha} \lvert  b_\text{ph}^\text{Sch}) & = -\dfrac{\left(b_M \left(\boldsymbol{\alpha}\right) - b_\text{ph}^\text{Sch}\right)^2}{2\sigma^2}.
\end{align}
For flat priors, the Fisher matrix \eqref{eq: FisherMatrix} becomes
\begin{align}
    F_{ij} \propto \, \delta g^{(i)}_{tt} \left(r_\text{ph}^\text{Sch}\right)\delta g_{tt}^{(j)} \left(r_\text{ph}^\text{Sch}\right),
\end{align}
which has only one non-zero eigenvalue, corresponding to the eigenvector 
\begin{align}
\boldsymbol{e}^{(1)} \propto  \left(\delta g_{tt}^{(1)}, \dots, \delta g_{tt}^{(M)}\right)\rvert_{r_\text{ph}^\text{Sch} }.
\end{align}
Then, the only combination [c.f. Eq.~(11) of Ref.~\cite{Volkel:2020xlc}] we can constrain with the EHT measurement is 
\begin{align}
    \left \lvert  \sum_{i = 1}^{M} \alpha_i \delta g_{tt}^{(i)} \left(r_\text{ph}^\text{Sch}\right)  \right \rvert  \lesssim  0.17 .
\end{align}

\bibliography{literature}
\end{document}